\documentclass[a4paper]{PoS}

\newcommand{\Xmaxmu}{X_{\mathrm{max}}^{\mu}}

\title{A branching model for hadronic air showers}

\ShortTitle{A branching model for hadronic air showers}

\author{\speaker{Vladimir Novotny}\\
        Charles University, Faculty of Mathematics and Physics, Institute of Particle and Nuclear Physics,
Prague, Czech Republic\\
        E-mail: \email{novotnyv@ipnp.troja.mff.cuni.cz}}        

\author{Dalibor Nosek\\
        Charles University, Faculty of Mathematics and Physics, Institute of Particle and Nuclear Physics,
Prague, Czech Republic\\
        E-mail: \email{nosek@ipnp.troja.mff.cuni.cz}}
        
\author{Jan Ebr\\
        Institute of Physics of the Academy of Sciences of the Czech Republic, Prague, Czech Republic\\
        E-mail: \email{ebr@fzu.cz}}

\abstract{We introduce a simple branching model for the development of hadronic showers in the Earth's atmosphere. 
Based on this model, we show how the size of the pionic component followed by muons can be estimated. 
Several aspects of the subsequent muonic component are also discussed. 
We focus on the energy evolution of the muon production depth. 
We also estimate the impact of the primary particle mass on the size of the hadronic component. 
Even though a precise calculation of the development of air showers must be left to complex Monte Carlo simulations, the proposed model can reveal qualitative insight into the air shower physics.}

\FullConference{The 34th International Cosmic Ray Conference,\\
		30 July- 6 August, 2015\\
		The Hague, The Netherlands}

\begin{document}

\section{Introduction}

We study a hadronic component of extensive air showers.
Key parameters that we want to determine are a shape of a muonic subshower profile and an atmospheric depth of its maximum ($\Xmaxmu$).
The $\Xmaxmu$ is the well known observable that can be used for a mass determination of primary cosmic ray particles, see e.g. Ref.~\cite{augerMPD} where detailed description of muon production depths reconstruction at the Pierre Auger Observatory together with mass composition implications of the measurement are summarized.

There are two methods that can be used for calculation of shower profiles.
The most precise of them is the use of Monte Carlo simulation tools that provide detailed information about showers.
A disadvantage of using Monte Carlo tools lays in the fact that they are computationally very demanding.
On the other hand, for a qualitative description and better physical insight one could use simple analytical or semi--analytical models that are in general easy to handle and have a clear interpretation.
We focus on the latter option.

The best known analytical model for an electromagnetic component of cosmic ray showers is the Heitler model \cite{heitler}. 
The Heitler model can be extended to the Heitler--Matthews model \cite{matthews} to incorporate hadronic (pionic) component.
The Heitler--Matthews model is further modified to the extended Heitler--Matthews model presented in Ref.~\cite{montanus2}.
The extended Heitler--Matthews model includes the better description of pion interaction lengths and branching multiplicities.
However, all of the above mentioned analytical models do not properly describe the shape of neither the electromagnetic nor the hadronic component.
The shape of the electromagnetic component can be well reproduced by the intermediate shower model that modifies the original Heitler model, as explained in Ref.~\cite{montanus1}.
In this study, we perform a similar alteration for the extended Heitler--Matthews model.
Because of the complexity of the hadronic component we use a semi--analytical approach based on the branching probabilities and a simple numerical integration.

\section{Method}

\begin{figure}
\begin{center}
\includegraphics[width=0.5\columnwidth]{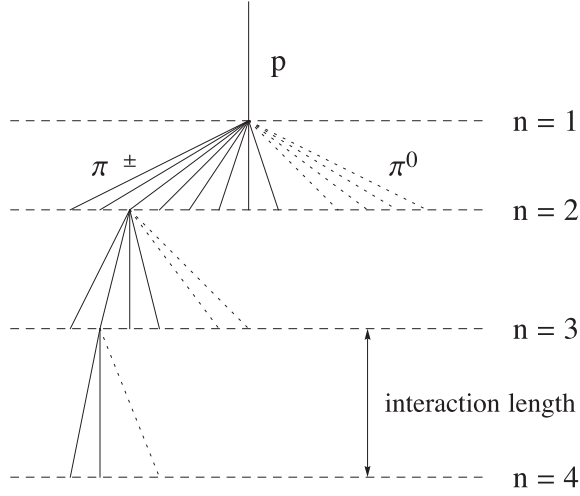}
\end{center}
\caption{The hadronic cascade for energy dependent interaction lengths in the extended Heitler--Matthews model. Figure is taken from Ref.~\cite{montanus2}.}
\label{fig:branching}
\end{figure}

For the description of the hadronic component we adopt the extended Heitler--Matthews model, see Ref.~\cite{montanus2}.
A basic idea of the extended Heitler--Matthews model is visualized in Fig.\ref{fig:branching}.
A primary particle interacts in the atmosphere and produces secondary charged and neutral pions.
The hadronic component within the model consists only of pions.
Neutral pions immediately split into two photons and produce the electromagnetic showers that we do not examine.
Charged pions further interact in the atmosphere and create other pions.
A multiplicity $M_{\pi}$ of produced pions in the pion initiated interaction in the shower is energy dependent \cite{montanus2}
\begin{equation}
M_{\pi} \approx 0.15 \cdot 	E^{0.18}.
\label{eq:M}
\end{equation}
$E$ is the energy of the initial pion.
We assume the ratio between charged and neutral pions to be $2:1$ in all interactions.
Because of that the multiplicity of charged pions is related to the total pion multiplicity by
\begin{equation}
M_{\pi^{+-}} = \frac{2}{3} M_{\pi}.
\end{equation}

The pion multiplicity of the first interaction depends on the mass of the primary particle.
It differs for primary protons and iron nuclei.
The pion multiplicity of the proton interaction is assumed to be equal to the pion multiplicity of the pion initiated interaction in the shower
\begin{equation}
M_{\mathrm{p}} \approx M_{\pi} = 0.15 \cdot E_0^{0.18}.
\end{equation}
For an iron primary it is considered to be
\begin{equation}
M_{\mathrm{Fe}} \approx 0.45 \cdot E_0^{0.18}
\end{equation}
where $E_0$ is the energy of the primary cosmic ray particle.

The most important simplification of the model is the presumption that the energy is equally divided among all produced particles in the interaction.
Therefore the pion energy after $n$ interactions (energy branching) is given by
\begin{equation}
E_{\pi,n} =\frac{E_0}{M_{\mathrm{p}/\mathrm{Fe}} \cdot M_{\pi}^{n-1}}.
\label{eq:E}
\end{equation}
Other necessary parameters of the model are the interaction lengths.
For pions, protons and iron nuclei we adopt \cite{montanus2}
\begin{eqnarray}
\lambda_{\pi-\mathrm{air}} [\mathrm{g~cm}^{-2}]. &\approx & 200 - 3.3 \log{E[\mathrm{eV}]}, \\
\lambda_{\mathrm{p}-\mathrm{air}} [\mathrm{g~cm}^{-2}] &\approx & 145 - 2.3 \log{E_0[\mathrm{eV}]}, \\
\lambda_{\mathrm{Fe}-\mathrm{air}} [\mathrm{g~cm}^{-2}] &\approx & 12.
\end{eqnarray}

In the extended Heitler--Matthews model \cite{montanus2}, pions in a shower produce a cascade until they reach a critical energy and then decay.
We do not use this assumption.
We presume that interaction and decay processes of pions compete.
To this end, we assign probabilities for the pion decay, for their interaction with the air nuclei or their survival.
When they interact they produce $M_{\pi}$ pions (Eq.\ref{eq:M}) with energies of $E_{\pi,n}$ (see Eq.\ref{eq:E}).
These assumptions allow us to calculate the number of pion decays in different atmosphere depths.
Muons are produced at the same atmospheric depths where pions decay.

In calculations we use an approximate relation between the depth in the atmosphere $X$ and height $h$
\begin{equation}
X(h) \mathrm{[g~cm^{-2}]} = 1030 \cdot \exp \left( {-\frac{h \mathrm{[km]}}{8}} \right).
\end{equation} 
We model the atmosphere as $N$ slant depths separated by a slant depth $\Delta X$.
We start from $N=0$ where the primary particle interacts and calculate the number of secondary pions and their energies.
Then we estimate the probabilities that pions decay, interact or survive until they reach the $N=1$ atmosphere depth.
We assume that corresponding fractions of pions undergo the processes.
Repeating all the calculations for the next $N$, and so on until the ground, gives us the energy distribution of pions in the atmosphere.
Only several energy bins are present because the energy is equally divided in interactions.
Since the interaction and decay probabilities depend on the depth in the atmosphere\footnote{The zenith angle of the incident particle has to be taken into account.} and on the pion energy, we calculate only several probability values for each atmospheric depth.

To reproduce the muon production depths (MPD) we have to incorporate the muon decay.
In our model, as it is in relevant experiments \cite{augerMPD}, these production depths are determined for muons that reach the ground.
Thus, for each muon, we calculate the probability that it arrives at the ground and assume this probability gives us the fraction of muons that contributes to the MPD.
Energy losses of muons in the air are neglected.

\section{Results}

\begin{figure}
\begin{center}
\includegraphics[width=0.8\columnwidth]{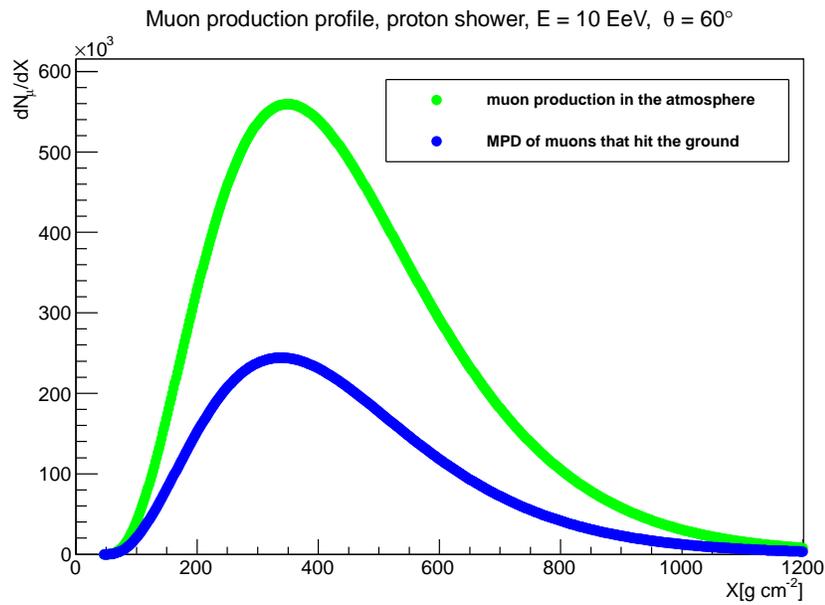}
\end{center}
\caption{Distribution of the production depth of muons. 
Green and blue curves represent respectively all produced muons and those that reach the ground.
Results are for a primary proton with energy of 10~EeV and incident zenith angle of 60$^\circ$.}
\label{fig:MPD_atm}
\end{figure}

The comparison between the depths of muon production in the atmosphere and production depths of muons that reach the ground is depicted in Fig.\ref{fig:MPD_atm}.
As shown, not only the total number of muons but also the shape of the profile and its maximum are affected by the muon decay in the atmosphere.
We have to include the decay process to get reliable results.

\begin{figure}
\begin{center}
\includegraphics[width=0.8\columnwidth]{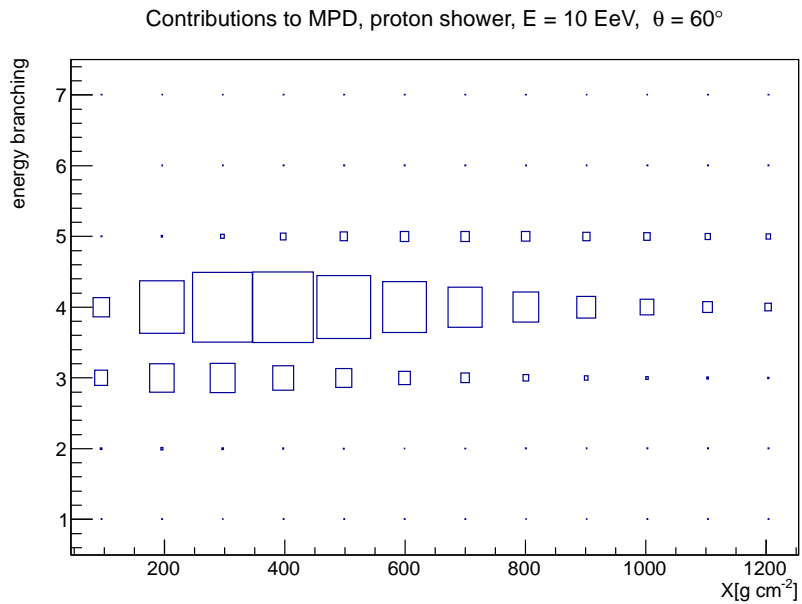}
\end{center}
\caption{
Contributions of pion decays at different energy branchings to the MPD for the example given in Fig.2.
}
\label{fig:energy_branching}
\end{figure}

The distribution of energy branchings at which the pion decay contributes to the MPD is shown in Fig.\ref{fig:energy_branching}.
It is clear that for the example given in Fig.\ref{fig:energy_branching} the majority of muons is generated at the 4$^{\mathrm{th}}$ branching but the 3$^{\mathrm{rd}}$ and 5$^{\mathrm{th}}$ can not be neglected, otherwise the shape of the MPD profile will be affected.
 
\begin{figure}
\begin{center}
\includegraphics[width=0.8\columnwidth]{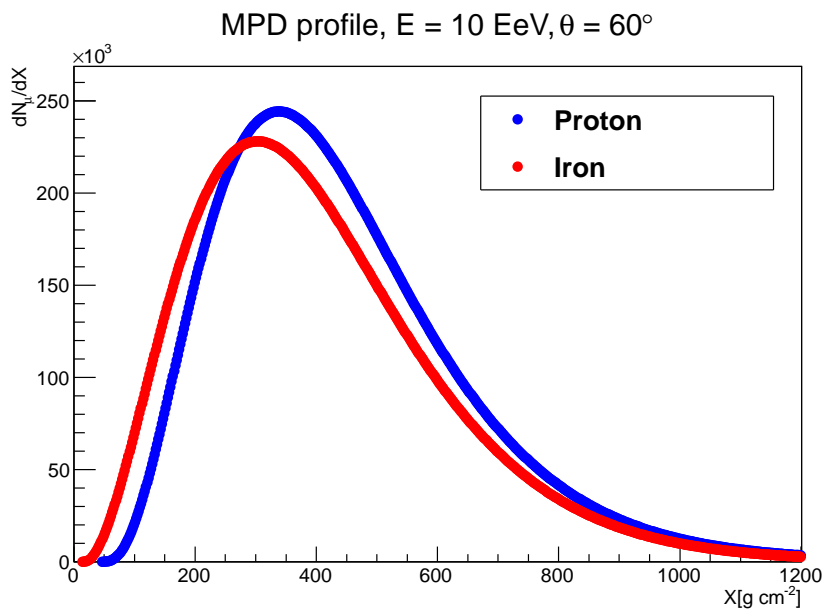}
\end{center}
\caption{MPD profiles for primary proton and iron particles with incident zenith angle of 60$^\circ$ at energy of 10 EeV.}
\label{fig:profile}
\end{figure}

The most important results are MPD profiles calculated for different primaries. 
As an example, in Fig.\ref{fig:profile} MPD profiles for proton and iron primaries with energy of 10~EeV are presented.
The maximum of the profile at the $\Xmaxmu$ for the proton 
induced shower is larger and occurs deeper in the atmosphere than the 
maximum of the profile for the primary iron.
Profiles in general correspond to those presented in Ref.~\cite{augerMPD}.

\begin{figure}
\begin{center}
\includegraphics[width=0.8\columnwidth]{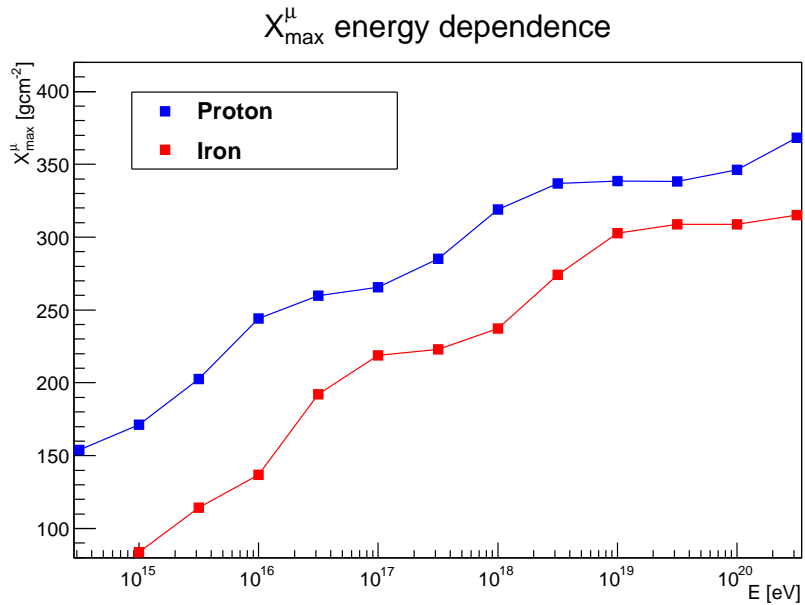}
\end{center}
\caption{Energy dependence of the $\Xmaxmu$ for protons and iron nuclei. Incident angles of the primaries are 60$^\circ$.}
\label{fig:Xmaxmu}
\end{figure}

The energy evolution of the $\Xmaxmu$ is shown in Fig.\ref{fig:Xmaxmu}.
The difference between proton and iron showers is clearly visible and the general trend of the increasing $\Xmaxmu$ as a function of energy is present.
As a result of muon decays in the atmosphere, we can see unexpected features in this dependence.
It is because muons are considered to have discrete energies at the production.
Different energies contribute differently to the MPD as is visualized in Fig.\ref{fig:energy_branching}.
The muon decay results in a suppression of lower energy branchings which leads to the decreasing of the $\Xmaxmu$ when the higher energy branching starts to dominate the MPD.

\section{Conclusions}

We utilized the idea of the extended Heitler--Matthews model \cite{montanus2, montanus1} to calculate the muon production in the extended air showers.
Despite the simplicity of the model we were able to reproduce MPD profiles for proton and iron primaries.
Values of $\Xmaxmu$ increase as a function of energy as expected but provide a systematic underestimation.
The flattenings in the $\Xmaxmu$ energy dependence show us the limits of the description and suggest that a more realistic energy division in the interactions is necessary for precise predictions of the $\Xmaxmu$.
Details of the MPD calculations in real experiments are complicated and there is no straightforward way to directly compare them with our results.

\acknowledgments
This work was supported by the Czech Science Foundation grant 14-17501S.


\begin{thebibliography}{99}

\bibitem{augerMPD} 
The Pierre Auger Collaboration, 
\emph{Muons in air showers at the Pierre Auger Observatory: Measurement of atmospheric production depth},
\emph{Physical Review D} {\bf 90}, 012012 (2014)
[{\tt hep-ex/1407.5919}].

\bibitem{heitler}
W. Heitler,
\emph{The Quantum Theory of Radiation}, Oxford Univ.
Press, London, 1954.

\bibitem{matthews}
J. Matthews,
\emph{A Heitler model of extensive air showers},
\emph{Astropart. Phys.} {\bf 22}, 387 (2005).

\bibitem{montanus2}
J. M. C. Montanus,
\emph{An extended Heitler-Matthews model for the full hadronic cascade in cosmic air showers},
\emph{Astropart. Phys.} {\bf 59}, 4 (2014)
[{\tt astro-ph.HE/1311.0642v3}].

\bibitem{montanus1} 
J. M. C. Montanus,
\emph{Intermediate models for longitudinal profiles of cosmic showers},
\emph{Astropart. Phys.} {\bf 35}, 651 (2012)
[{\tt astro-ph.HE/1106.1073v2}].

\end{thebibliography}
\end{document}